\documentclass[aps,prb,nobibnotes,twocolumn,superscriptaddress,bibliography]{revtex4-2}
\usepackage{amsfonts}
\usepackage{mathrsfs}
\usepackage{amsmath}
\usepackage{color}
\usepackage{graphicx}
\usepackage{bm}
\usepackage{amssymb}
\usepackage{xspace}
\usepackage{epstopdf}
\usepackage{dcolumn}
\usepackage{longtable}
\usepackage{multirow}
\usepackage{float}
\usepackage{comment}
\usepackage{pifont}
\usepackage{bbding}
\usepackage{fontawesome}
\usepackage{soul}

\usepackage[colorlinks=true, letterpaper=true, pdfstartview=FitV, linkcolor=blue, citecolor=blue, urlcolor=blue]{hyperref}

\makeatother

\begin{document}
\title{Electric Hall Effect and Quantum Electric Hall Effect}
\author{Chaoxi Cui}
\affiliation{Centre for Quantum Physics, Key Laboratory of Advanced Optoelectronic
Quantum Architecture and Measurement (MOE), School of Physics, Beijing
Institute of Technology, Beijing 100081, China}
\affiliation{Beijing Key Lab of Nanophotonics \& Ultrafine Optoelectronic Systems,
School of Physics, Beijing Institute of Technology, Beijing 100081,
China}

\author{Run-Wu Zhang}
\email{zhangrunwu@bit.edu.cn}
\affiliation{Centre for Quantum Physics, Key Laboratory of Advanced Optoelectronic
Quantum Architecture and Measurement (MOE), School of Physics, Beijing
Institute of Technology, Beijing 100081, China}
\affiliation{Beijing Key Lab of Nanophotonics \& Ultrafine Optoelectronic Systems,
School of Physics, Beijing Institute of Technology, Beijing 100081,
China}

\author{Yilin Han}
\affiliation{Centre for Quantum Physics, Key Laboratory of Advanced Optoelectronic
Quantum Architecture and Measurement (MOE), School of Physics, Beijing
Institute of Technology, Beijing 100081, China}
\affiliation{Beijing Key Lab of Nanophotonics \& Ultrafine Optoelectronic Systems, School of Physics, Beijing Institute of Technology, Beijing 100081,
China}

\author{Zhi-Ming Yu}
\email{zhiming\_yu@bit.edu.cn}
\affiliation{Centre for Quantum Physics, Key Laboratory of Advanced Optoelectronic Quantum Architecture and Measurement (MOE), School of Physics, Beijing
Institute of Technology, Beijing 100081, China}
\affiliation{Beijing Key Lab of Nanophotonics \& Ultrafine Optoelectronic Systems, School of Physics, Beijing Institute of Technology, Beijing 100081, China}

\author{Yugui Yao}
\email{ygyao@bit.edu.cn}
\affiliation{Centre for Quantum Physics, Key Laboratory of Advanced Optoelectronic Quantum Architecture and Measurement (MOE), School of Physics, Beijing
Institute of Technology, Beijing 100081, China}
\affiliation{Beijing Key Lab of Nanophotonics \& Ultrafine Optoelectronic Systems, School of Physics, Beijing Institute of Technology, Beijing 100081, China}

\begin{abstract}
Exploring new Hall effect is always a fascinating research topic. The ordinary Hall effect and the quantum Hall effect, initially discovered in two-dimensional (2D) non-magnetic systems, are the phenomena that  a transverse current is generated when a system carrying an electron current is placed in a magnetic field perpendicular to the currents.
In this work, we propose the electric counterparts of these two Hall effects, termed as electric Hall effect (EHE) and quantum electric Hall effect (QEHE). The EHE and QEHE emerge in 2D magnetic systems, where the transverse current is generated by applying an electric gate-field instead of a magnetic field.
We present a symmetry requirement for  intrinsic  EHE and QEHE.
With a weak gate-field, we establish an analytical expression of the intrinsic EHE coefficient. We show that it is determined by intrinsic band geometric quantities: Berry curvature and its  polarizability which  consists of both intraband and interband  layer polarization.
Via first-principles calculations, we  investigate the EHE in the monolayer Ca(FeN)$_2$, where significant EHE coefficient is observed around band crossings.
Furthermore, we demonstrate that the QEHE can appear in  the semiconductor  monolayer $\rm BaMn_2S_3$, of which the  Hall conductivity  exhibits steps that take on the quantized values  $0$ and $\pm1$ in the unit of $e^2/h$ by varying the gate-field within the experimentally achievable range.
Due to the great tunability of the electric gate-field, the EHE and QEHE proposed here can be easily controlled and should have more potential applications.
\end{abstract}
\maketitle
As a fundamental transport phenomenon, the Hall effect, discovered over a century ago, remains  an actively expanding area of condensed matter physics   \cite{HallEarliest,Hall1881,pugh1953hall,klitzing1980new,von1986quantized,yennie1987integral,murthy2003hamiltonian,nagaosa2010anomalous,sinova2015spin}.
Various types of Hall effects, including ordinary and planar (in-plane), intrinsic and extrinsic, linear and nonlinear Hall effects, continue to be proposed and investigated
  \cite{PlanarHallEarliest,NonLinearHallEarliest,NonlinearGaoYang,QuantumnonlinearHall,TaskinPlanarHall,NonlinearNagaosa,NonlinearplanarNagaosa,NonLinearHall,IntrinsicSecond-OrderAnomalousHall,IntrinsicNonLinearHallCuMnAs,IntrinsicNonlinearPlanarHall,LiLeiPlanar,chang2023colloquium,park2023observation,lyalin2023magneto,cao2023plane,wang2024orbital,wang2024orbital}.
The ordinary Hall effect occurs in systems with crossed electric and magnetic fields, where the transverse current is perpendicular to both fields.
Since the transverse current is generated by the magnetic field ($\mathcal{B}$-field), the Hall conductivity vanishes when the strength of the $\mathcal{B}$-field approaches zero.
Moreover, when the  $\mathcal{B}$-field is large enough, the Hall conductivity of a metal can be quantized, leading to the quantum Hall effect  \cite{klitzing1980new,laughlin1981quantized,aoki1981effect,cage2012quantum}.

\begin{table*}[t]
	\caption{Symmetry requirements of the intrinsic  EHE. ``\ding{51}" and  ``\ding{55}" denote that the element is symmetry allowed and forbidden, respectively. The magnetic layer-point-group symmetry operators are divided into four parts: ${\cal{S}}$,  $\mathcal{T} {\cal{S}}$, $M_z {\cal{S}}$ and  $M_z \mathcal{T} {\cal{S}}$.
Notice that $S_{2,z}={\cal{P}}$.
}.
	\begin{ruledtabular} %
\begin{tabular}{ccccc}
	   & ${\cal{S}}$ & $\mathcal{T} {\cal{S}}$ & $M_z {\cal{S}}$ & $M_z \mathcal{T} {\cal{S}}$ \\
	  Elements & $\left\{E, M_{\|} \mathcal{T}, C_{n,z}\right\}$ & $\left\{\mathcal{T}, M_{\|}, C_{n,z} \mathcal{T}\right\}$ & $\left\{M_z, C_{2,\|} \mathcal{T}, S_{n,z}\right\}$ & $\left\{M_z \mathcal{T}, C_{2,\|}, S_{n,z} \mathcal{T}\right\}$ \\
	
	$\sigma_{a b}$ & \ding{51} &  \ding{55} & \ding{51} &  \ding{55} \\
	$\chi_{a b}$ & \ding{51} &  \ding{55} &  \ding{55} & \ding{51} \\
	 Requirement & optional & forbidden & forbidden & requisite \\
\end{tabular}
	\end{ruledtabular}
	\label{table1}
\end{table*}

In addition to the $\mathcal{B}$-field, the electric gate-field ($\mathcal{E}$-field) is another common external field used to manipulate the physical properties of systems   \cite{ElectricControlFerromagnetism,ElectricControlSpinTransport,ElectricControlSpin,chun2012electric,heron2014deterministic,noel2020non,gao2020tunable,LayerHall}. Particularly in 2D materials, the $\mathcal{E}$-field is directly coupled  to the layer polarization of electron bands, providing an efficient and predictable method to control  the band structure and many physical observations of the systems   \cite{ValleyLayerCouping,SpinValleyLayer,Cornertronics,TiSiCOTB}. Furthermore, electric control of spin and valley polarization has been extensively investigated in both magnetic and non-magnetic systems \cite{wolf2001spintronics,bader2010spintronics,qian2014quantum,manchon2015new,schaibley2016valleytronics,vitale2018valleytronics,Marfoua_2020,qin2020noncollinear,zheng2022coupling,liang2023tunable}.
However, the use of a static $\mathcal{E}$-field instead of a $\mathcal{B}$-field to generate  a net Hall effect and the quantum Hall effect  has not been well  investigated.

\begin{figure}
	\includegraphics[width=8. cm]{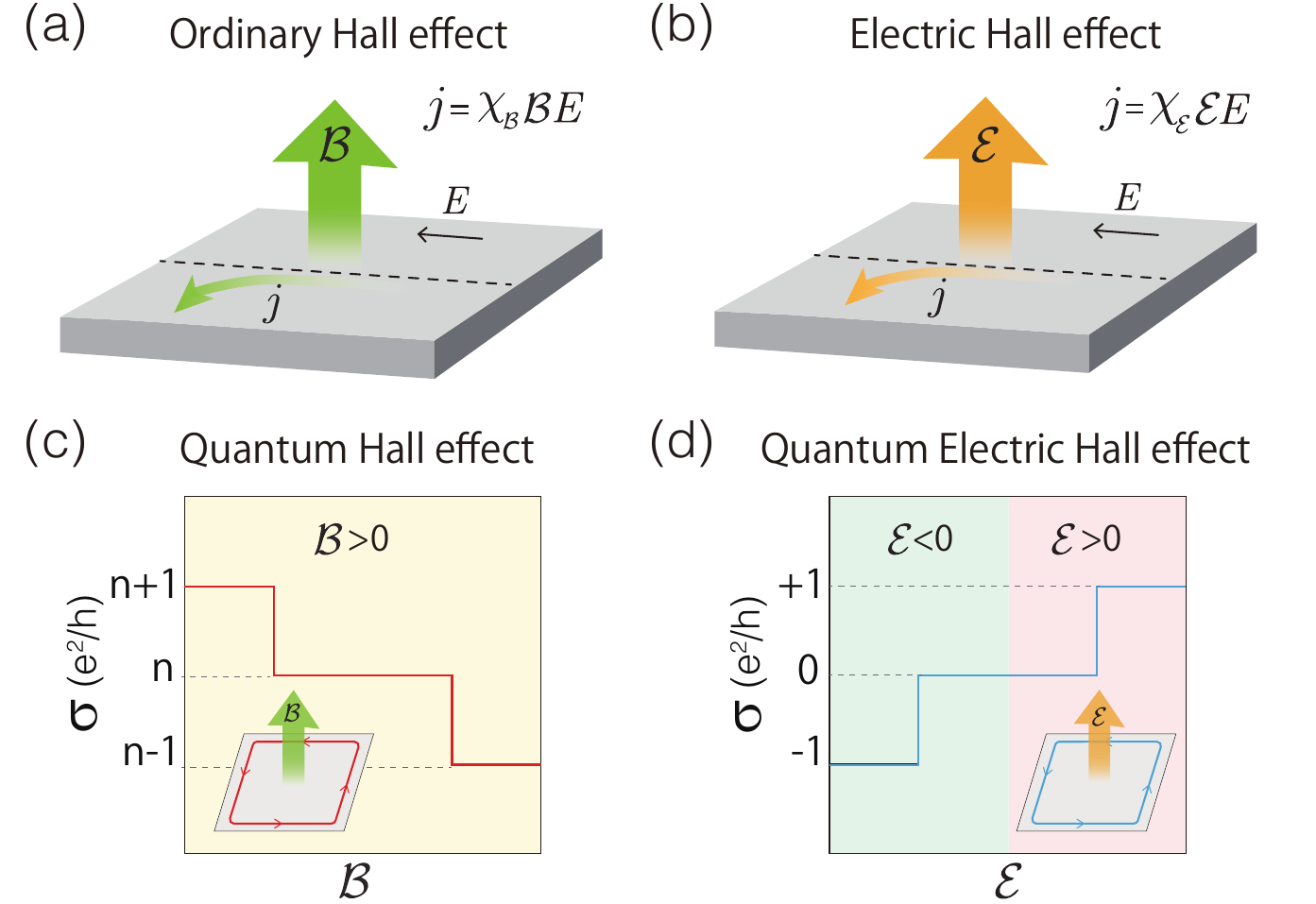}
	\caption{Schematics for (a) ordinary  and (c) quantum Hall effects. 
(b) and (d) illustrate EHE and QEHE, which can be considered as the  electric counterparts of  ordinary  and  quantum Hall effects.}
	\label{Fig. 1}
\end{figure}

To address this important task,  we in this work explore the electric counterparts of the ordinary and quantum Hall effects in 2D magnetic systems, where the Hall current arises from a perpendicular static $\mathcal{E}$-field instead of $\mathcal{B}$-field, as illustrated in Fig.  \ref{Fig. 1}. Through symmetry analysis, we find that the intrinsic EHE can   exist  in  a wide range of 2D magnetic systems,  including ferromagnetic, antiferromagnetic, and altermagnetic materials.
We  derive the   analytical expression of    the  intrinsic EHE, and show that it  is closely related to  Berry curvature and its polarizability consisting of both   intraband and interband  layer polarization.
This means that the EHE will be pronounced around band degeneracies  and the band edges of systems with small band gaps.
We validate the existence  of EHE by  first-principle calculations on  monolayer Ca(FeN)$_2$, where significant  EHE is observed when the Fermi energy is around the  Weyl points.
Furthermore, we predict that an experimentally accessible $\mathcal{E}$-field  can induce QEHE in monolayer  $\rm BaMn_2S_3$.
Remarkably, the quantized Hall conductivity can be  switched by inverting the $\mathcal{E}$-field.
Therefore, our work not only proposes a new mechanism for the Hall effect but also offers an efficient and convenient method for generating and controlling it.

\textcolor{blue}{\emph{Intrinsic EHE and general analysis.–}}
Consider a 2D system within the $x$-$y$ plane.
The  setup for the EHE proposed here is similar to that of the ordinary Hall effect, as illustrated in Fig.  \ref{Fig. 1}(b).
For EHE (ordinary Hall effect), the applied electric gate-field ${\mathcal{E}}_z$ (magnetic field ${\mathcal{B}}_z$) is perpendicular to the sample, and the transverse response current ${\bm j}$ is normal to the driving ${\bm E}$ field.
The intrinsic EHE coefficient $\chi_{ab}$ (irrelevant to  relaxation time  $\tau$) to the leading order can be expressed as
\begin{equation}
	j_{a}=\chi_{ab}{\mathcal{E}}_zE_{b}=\sigma_{ab}E_b, \label{Eq1}
\end{equation}
where  $\sigma_{ab}=\chi_{ab}{\mathcal{E}}_z$ denotes the electric Hall conductivity, and the indices  $\{a,b\}\in\{x,y\}$ with $a\ne b$. Notice that ${\mathcal{E}}_z$ and  ${\bm E}$  are two \emph{independent} electric fields. 
Similar to ordinary Hall effect,   the  Hall current in EHE vanishes when  ${\mathcal{E}}_z=0$.
However,  the symmetry conditions for the  existence of EHE and the ordinary Hall  are  completely different, due to the distinct symmetry constraints on ${\mathcal{E}}_z$ and ${\mathcal{B}}_z$.

From Eq. (\ref{Eq1}), one  knows that for  a system exhibiting EHE, it should not have a symmetry operator that leads to  $\chi_{ab}=0$, but must exhibit  at least one operator that enforces $\sigma_{ab}=0$ in the absence of ${\mathcal{E}}_z$.
Since the difference between  $\chi_{ab}$ and $\sigma_{ab}$ is ${\mathcal{E}}_z$, and the mirror  symmetry  $M_z$ can reverse  ${\mathcal{E}}_z$ while keeps the driving ${\bm E}$ field and $\bm j$ unchanged, we can decompose the  magnetic layer-point-group symmetry operators  into two parts: ${\cal{O}}$ and $M_z{\cal{O}}$, where ${{\cal{O}}=\{E,C_{n,z},M_{\|}\}\otimes\{E,\mathcal{T}\}}$ (with $n=2,3,4,6$) preserves ${\mathcal{E}}_z$ and $M_z{\cal{O}}$ reverse ${\mathcal{E}}_z$.
Here, $E$ represents identity element and  $M_{\|}$ denotes any  mirror operator perpendicular to the plane.
Under this decomposition,  it is obvious that the operators in ${\cal{O}}$ allows or annihilate both $\sigma_{ab}$ and $\chi_{ab}$.
In contrast, an operator in $M_z{\cal{O}}$ that allows (annihilates) $\sigma_{ab}$ will annihilate (allow) $\chi_{ab}$. 
As $\mathcal{T}$ can reverse $\bm j$ while keeps $\bm E$ and ${\mathcal{E}}_z$ unchanged, ${\cal{O}}$ can be further subdivided into  ${\cal{O}}={\cal{S}}+{\cal{TS}}$, where ${\cal{S}}=\{E,M_{\|} \mathcal{T},C_{n,z}\}$ allows  $\sigma_{ab}$ and $\chi_{ab}$, while ${\cal{TS}}=\{\mathcal{T},M_{\|},C_{n,z}\mathcal{T}\}$ annihilates them.
Accordingly, $M_z{\cal{O}}$  is  separated  into $M_z{\cal{O}}=M_z{\cal{S}}+M_z{\cal{TS}}$, where $M_z{\cal{S}}=\{M_z,C_{2,\|}\mathcal{T},S_{n,z}\}$ and $M_z\mathcal{T}{\cal{S}}=\{M_z\mathcal{T},C_{2,\|},S_{n,z}\mathcal{T}\}$ with  $C_{2,\|}$ denoting  any in-plane two-fold rotation. 
As earlier discussed, since  ${\cal{S}}$ permits both  $\sigma_{ab}$ and $\chi_{ab}$, $M_z{\cal{S}}$ will also allows $\sigma_{ab}$ but annihilates $\chi_{ab}$. Similarly,  $M_z\mathcal{T}{\cal{S}}$ annihilates $\sigma_{ab}$ but allows $\chi_{ab}$.
Consequently, for  a system hosting finite  EHE,  it should not have  $\mathcal{T}{\cal{S}}$ and $M_z{\cal{S}}$ as  they can annihilates $\chi_{ab}$, while it  must have at least one symmetry in $M_z\mathcal{T}{\cal{S}}$ to annihilates $\sigma_{ab}$. The symmetries in ${\cal{S}}$ are  optional, as they  have no influence on the existence of  $\sigma_{ab}$ and $\chi_{ab}$.

The results of the symmetry analysis are  summarized in Table \ref{table1}, showing that the EHE can manifest  in a wide range of magnetic materials.
For example, the  ${\cal{PT}}(=S_{2,z}{\cal{T}})$  symmetry  in  $M_z\mathcal{T}{\cal{S}}$  indicates that the EHE can appear in antiferromagnetic materials.
Besides,  $C_{2,\|}$ can be the symmetry operator of  a ferromagnetic material  with  in-plane spin polarization, and that of an altermagnetic material with out-of-plane N\'eel vector.

Furthermore, the magnetic materials with negligible spin-orbit coupling (SOC) respect the spin group symmetry \cite{brinkman1966theory,litvin1974spin,liu2022spin,jiang2023enumeration,ren2023enumeration,xiao2023spin}, and  exhibit some  emergent symmetries that are  somehow counterintuitive.
The most important one is the effective  $\cal{T}$ symmetry, which surprisingly exists in the collinear  or coplanar magnetic systems materials without SOC   \cite{SM}. 
This effective  $\cal{T}$ symmetry will eliminate both anomalous Hall effect and EHE. This means that the EHE may be  sensitive to  the SOC strength of the systems, as it can be realized in the collinear or coplanar  magnetic materials with strong  SOC but disappears when  SOC is absent.

\textcolor{blue}{\emph{Expression of intrinsic $\chi_{ab}$.–}} \label{Sec2}
The expression of the  intrinsic EHE coefficient  $\chi_{ab}$ can be established by perturbation theory. 
Without ${\mathcal{E}}_z$ field, the intrinsic  Hall conductivity of the  system is zero.
When ${\mathcal{E}}_z$ is  finite but weak, the   Hamiltonian of the perturbed system  reads
\begin{equation}
	\hat{H}=\hat{H}_{0}+g_{E}\mathcal{E}_z\hat{z},
\end{equation}
with $\hat{H}_{0}$ the unperturbed Hamiltonian, $\hat{z}$ the position operator, and $g_{E}$ the $g$-factor-like coefficient, which is affected by screening effect and can be determined by DFT calculation.
The intrinsic  Hall conductivity of the perturbed system is obtained from the  integral of Berry curvature of all occupied bands \cite{nagaosa2010anomalous,vanderbilt2018berry,chang2023colloquium}
\begin{equation}
	\sigma_{x y}=\frac{e^{2}}{\hbar} \sum_{n}\int \frac{d^2 k}{(2 \pi)^2} f(\tilde{\varepsilon}_{n,k})\tilde{\Omega}_n(\boldsymbol{k}),
	\label{eq2}
\end{equation}
where $f$ is the Fermi-Dirac distribution, $\tilde{\varepsilon}_{n}$ and $\tilde{\Omega}_{n}$  are the  perturbed  band dispersion   and   Berry curvature, respectively.
To the first order of ${\mathcal{E}}_z$, we have
\begin{eqnarray}
	\tilde{\varepsilon}_{n} (\boldsymbol{k}) & = & \varepsilon_{n} (\boldsymbol{k})+P_{nn}(\boldsymbol{k})\mathcal{E}_z, \label{e1} \\
	\tilde{\Omega}_{n}(\boldsymbol{k}) & = & \Omega_{n}(\boldsymbol{k})+\Lambda_n(\boldsymbol{k})\mathcal{E}_z, \label{e2}
\end{eqnarray}
where  $\varepsilon_{n}$ and $\Omega_{n}$ respectively are  the  dispersion  and Berry curvature of the original state $|u_{n}(\boldsymbol{k})\rangle$, $P_{nm}=g_E \langle u_{n}(\boldsymbol{k})|\hat{z}|u_{m}(\boldsymbol{k})\rangle$ denotes the effective  layer polarization matrix element, and $\Lambda_n(\boldsymbol{k})= \partial_{\mathcal{E}_z}\Omega_{n}(\boldsymbol{k})|_{\mathcal{E}_z=0}$ is the Berry curvature polarizability, expressed as
\begin{eqnarray}
  \Lambda_n(\boldsymbol{k}) &=& -2\textrm{Im}\sum_{m\neq n}\left[-2\frac{P_{mm}-P_{nn}}{\left(\varepsilon_{m}-\varepsilon_{n}\right)^{3}}v_{x}^{nm}v_{y}^{mn}\right. \nonumber\\
   && +\sum_{l\neq m}\frac{P_{lm}^{*}v_{y}^{ln}v_{x}^{nm}+P_{lm}v_{x}^{nl}v_{y}^{mn}}{\left(\varepsilon_{n}-\varepsilon_{m}\right)^{2}\left(\varepsilon_{m}-\varepsilon_{l}\right)} \nonumber\\
   && \left.+\sum_{l\neq n}\frac{P_{ln}v_{y}^{ml}v_{x}^{nm}+P_{ln}^{*}v_{x}^{lm}v_{y}^{mn}}{\left(\varepsilon_{n}-\varepsilon_{m}\right)^{2}\left(\varepsilon_{n}-\varepsilon_{l}\right)}\right],
\end{eqnarray}
with $v^{nm}_{x(y)}=\langle u_{n}(\boldsymbol{k})|\hat{v}_{x(y)}|u_{m}(\boldsymbol{k})\rangle$  the velocity matrix element.
Similarly,  one has $f(\tilde{\varepsilon}_{n})  =  f(\varepsilon_{n})+\frac{\partial f}{\partial\varepsilon_n}P_{nn}(\boldsymbol{k})\mathcal{E}_z$.
Substituting $f(\tilde{\varepsilon}_{n})$ and $\tilde{\Omega}_n(\boldsymbol{k})$ into Eq. (\ref{eq2}), the intrinsic EHE coefficient  $\chi_{xy}$ (up to the leading order) is obtained  as  \cite{SM}
\begin{eqnarray}\label{eq:chi}
  \chi_{x y} &=& \frac{e^2}{\hbar} \sum_n \int  \frac{d^2 k}{(2 \pi)^2} \nonumber\\
   && \times\left[\frac{\partial f(\varepsilon_n)}{\partial \varepsilon_n}  P_{nn}(\boldsymbol{k}) \Omega_n(\boldsymbol{k}) +f(\varepsilon_{n}) \Lambda_n(\boldsymbol{k})\right],
\end{eqnarray}
indicating that the intrinsic  EHE results  from the change of the band structure [Eq. (\ref{e1})] and the  Berry curvature  [Eq. (\ref{e2})] of the systems induced by $\mathcal{E}_z$.
Besides, we   identify the following important  features of $\chi_{ab}$.

First, the first term in Eq. (\ref{eq:chi}) is a Fermi surface property, but the second term is calculated by the integral of  all the occupied bands.
Both terms are solely determined by the  intrinsic band  quantities of systems, and  can thus be easily  implemented in the first-principles calculations.
Moreover, due to the presence of Berry curvature and its polarizability in Eq. (\ref{eq:chi}), the EHE would be pronounced   around band crossing points and the band edges of the systems with narrow band gap.

Second, while the  first term of  $\chi_{xy}$ is in direct proportion to the intraband layer polarization $P_{nn}$, the  second term  has contribution from interband layer polarization $P_{nm}$  ($n\neq m$), which can be finite even if  $P_{nn}=P_{mm}=0$.
 This means that the  appearance of EHE does not require the  bands around Fermi surface to be  (intraband) layer polarized.

Third, the fully occupied bands  do not make any contribution to $\chi_{xy}$, because the intrinsic Hall conductivity of fully occupied bands in a 2D system is a topological quantity, characterized by  Chern number, which is robust against small  perturbations \cite{chang2023colloquium}. 
Thus, the $\chi_{xy}$ discussed here should be zero for semiconductors and insulators.
However, when   ${\mathcal{E}}_z$ is  strong enough to cause band inversion, a QEHE may appear as discussed below.

At last, it would be instructive to discuss the relationship  between EHE and the recently proposed layer Hall effect \cite{gao2021layer,feng2023layer,zhang2023layer,PhysRevB.107.085411,chen2024layer}. 
Again, taking the ordinary Hall effect as a comparison, one can find that this relationship is  completely similar to that between ordinary Hall and spin Hall effects, as  ${\cal{E}}$-field and ${\cal{B}}$-field are respectively  coupled to layer  and spin degrees of freedom.
Thus, it is apparent that  the systems having  layered Hall effect would   have EHE, but not vice versa, which also can be inferred from the  second feature of  $\chi_{xy}$.
Besides, the EHE may appear in  2D buckled systems, as their band and  Berry curvature can be affected by ${\cal{E}}$-field. 
However, for such systems it is not meaningful  to introduce the layer degree of freedom, and then the layer Hall effect.

\begin{figure}[t]
	\includegraphics[width=8.8 cm]{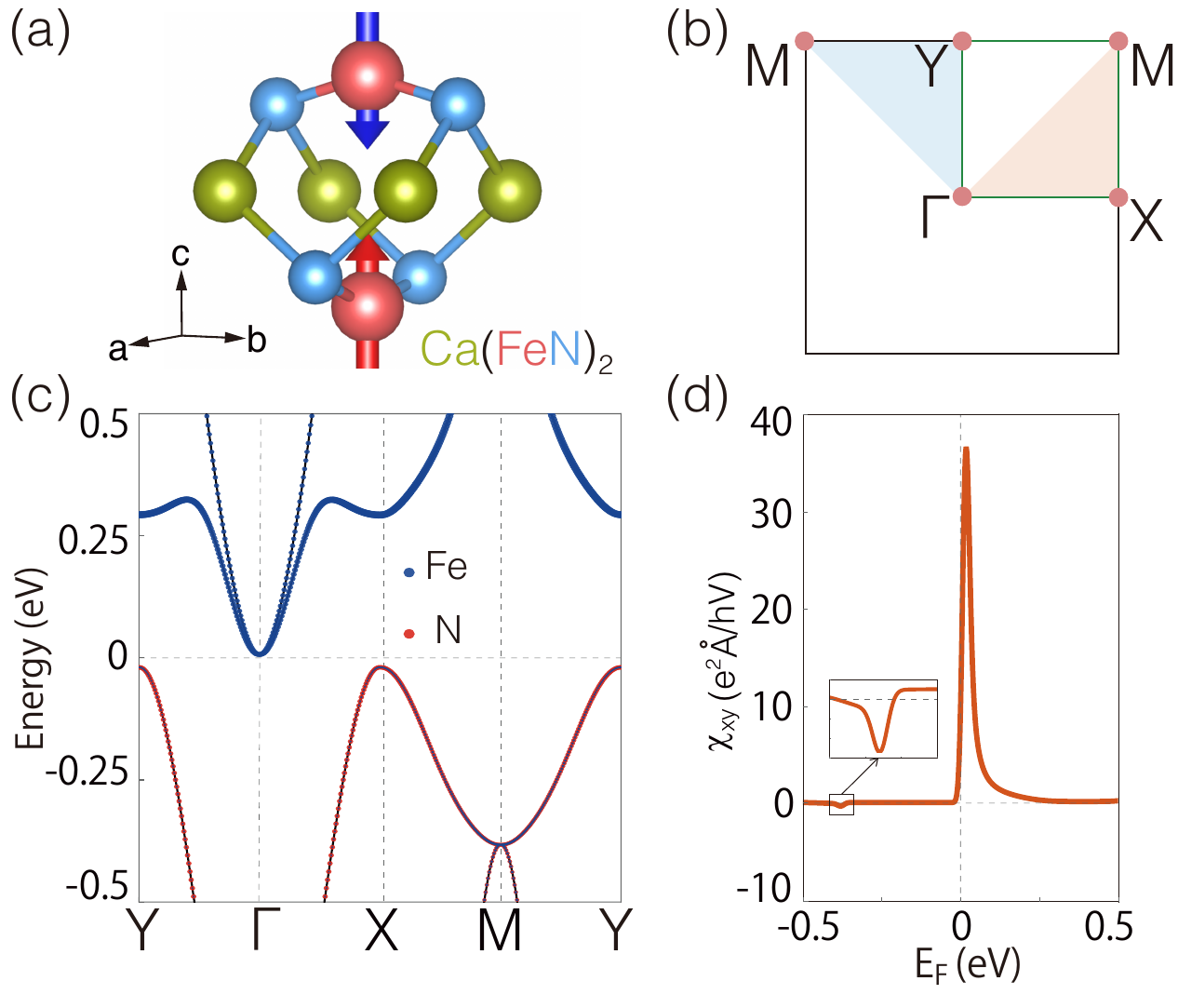}
	\caption{(a) Structure and (b) BZ of monolayer $\rm Ca(FeN)_2$. (c)  Orbital-projected  band structure   of $\rm Ca(FeN)_2$ with SOC. (d) $\chi_{xy}$  as a function of $E_F$. The inset shows a  small peak resulting  from the LWP at M point. Here  $g_E=0.23$ is used, which is obtained from DFT calculations. 
}
	\label{Fig. 2}
\end{figure}

\textcolor{blue}{\emph{EHE and material candidate.–}}
Guided by symmetry analysis  and the expression of EHE coefficient, we propose monolayer $\rm Ca(FeN)_2$ as a material  candidate hosting EHE.
Monolayer $\rm Ca(FeN)_2$  has a square lattice structure with optimized lattice constant $a=b=3.65$ \AA~\cite{SM}, as shown in Fig.  \ref{Fig. 2}(a). 
Its ground state  exhibits a   collinear altermagnetic configuration with  N\'eel vector pointing out of the plane, and the magnetic moments are mainly on the two inequivalent Fe sites with a magnitude about $ \pm3.54 \mu_B$.
Importantly, monolayer $\rm Ca(FeN)_2$ belongs to magnetic layer group  (MLG) No. 414 (59.5.414), which  has $S_{4z}{\cal{T}}$ symmetry satisfying the symmetry requirement listed in Table \ref{table1}. Moreover, the  SOC effect in  $\rm Ca(FeN)_2$ is not negligible.



The electronic band structure of the monolayer $\rm Ca(FeN)_2$ with SOC is plotted in  Fig.  \ref{Fig. 2}(c), where the orbital component of the electronic states  is also  presented.
A key observation is that there exist two band crossings at $\Gamma$  and  $M$ points, respectively.
Since both  $\Gamma$  and  $M$ points of MLG No. 414 have only one 2D irreducible band representation, it is easy to know that the two  band crossing are linear Weyl points (LWPs)  \cite{zhang2023encyclopedia}.
One can expect that the intrinsic EHE coefficient $\chi_{xy}$ would be sizable  around these two LWPs.

Based on the first-principle calculations and Eq. (\ref{eq:chi}), we calculate $\chi_{xy}$ of the monolayer $\rm Ca(FeN)_2$ as a function of Fermi energy $E_F$.
The result is shown in Fig.  \ref{Fig. 2}(d), in which two  peaks can be found around $0$ and $-0.38$ eV, which are exactly the energy position of the two LWPs  at $\Gamma$ and $M$ points.
As discussed above, for the collinear magnetic systems, the EHE has a strong dependence on the SOC strength, as it will vanish when SOC is absent.
Here,  the LWP at $\Gamma$  ($M$) point is mainly composed of  the $\rm Fe$  ($\rm N$) atoms [see Fig.  \ref{Fig. 2}(c)].
The SOC  strength of Fe atom is much stronger than that of N atoms.
Thus,   the peak from the LWP at $\Gamma$ point is  much more pronounced than that of $M$ point, as shown in  Fig.  \ref{Fig. 2}(d).
Besides, due to the large local band gap at $X$ and $Y$ points, $\chi_{xy}$ is rather small at the valence band edge.

From the calculations of the intrinsic EHE coefficient, one knows that  under a ${\cal{E}}_z$ field of $0.01$ eV/\AA, the intrinsic electric Hall conductivity can reach $\sim 0.3~e^2/h$ for lightly electron doping. This value is comparable to the Hall conductivity in many 2D materials, and is detectable  in experiment \cite{Fox2017PartpermillionQA,wang2020topological,gao2021layer,xu2023observation}.

\textcolor{blue}{\emph{QEHE and material candidate.–}}
Similar to the ordinary Hall effect, the EHE can also have its  quantized version.
The quantum Hall effect appears in 2D  systems that are subjected to  strong ${\cal{B}}$ field  \cite{von1986quantized,yennie1987integral}.
However, with suitable 2D materials, the QEHE can occur with moderate ${\cal{E}}_z$ field.

\begin{figure}[t]
	\includegraphics[width=8.8 cm]{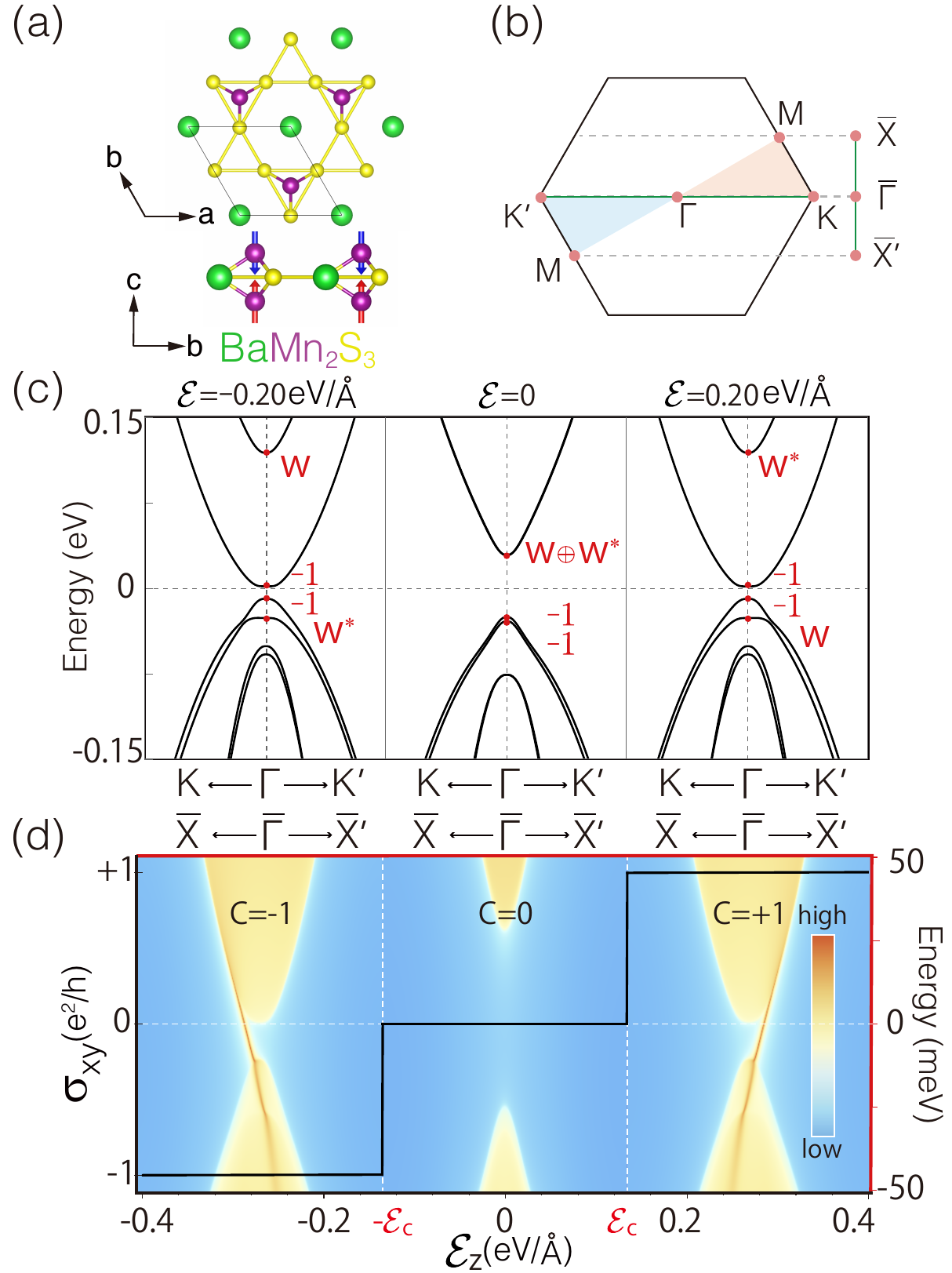}
	\caption{ (a) Structure and (b) BZ of monolayer $\rm BaMn_2S_3$. (c) Band structure of $\rm BaMn_2S_3$ with SOC for  ${\cal{E}}_z=0, \pm0.2$ eV/\AA. $C_{3,z}$ eigenvalue of states at  $\Gamma$ point is highlighted  in red with $W=e^{-i\pi/3}$. (d) Intrinsic Hall conductivity $\sigma_{ab}$ (left axis) of  $\rm BaMn_2S_3$ as a function of  ${\cal{E}}_z$ (down axis). The (10) edge states for   ${\cal{E}}_z=0, \pm0.2$ eV/\AA~(up and right axis)  is plotted as background. }
	\label{Fig. 3}
\end{figure}

To directly demonstrate it, we search for the materials having narrow direct band gap and at least one symmetry operator in  $M_z{\cal{TS}}$ (see Table \ref{table1}), as the operator in  $M_z{\cal{TS}}$ guarantees $\sigma_{ab}=0$ but is broken by ${\cal{E}}_z$ field.
We find that the monolayer $\rm BaMn_2S_3$  \cite{SM} belongs to  MLG No. 514 (78.5.514) and then  falls into this requirement.  
Monolayer $\rm BaMn_2S_3$  is structured in a triangular lattice with lattice constant $a=b=6.74$ \AA ~\cite{SM}.
The   magnetic configuration of monolayer $\rm BaMn_2S_3$  in ground state are depicted  in Fig.  \ref{Fig. 3} (a) with N\'eel vector oriented perpendicular to the plane. 
The magnetic moments primarily located on two  Mn atoms with the magnitude about  $ \pm 4.33 \mu_B$.
The monolayer has  $M_z\cal{T}$ symmetry, which forbids the intrinsic Hall conductivity but can be broken by ${\cal{E}}_z$. 
Besides, it  has $C_{3z}$ rotation symmetry, which still preserves when ${\cal{E}}_z\neq0$.
The band structure of monolayer $\rm BaMn_2S_3$ with SOC is plotted in the middle figure of  Fig.  \ref{Fig. 3}(c), showing it is a semiconductor with a direct gap of $54.5$ meV at $\Gamma$ point.

By  applying ${\cal{E}}_z$ field and increasing its strength, the band gap of system  becomes small and closes at a  moderate value  ${\cal{E}}_z=\pm {\cal{E}}_c$ with  ${\cal{E}}_c=0.135$ eV/\AA, which is  experimentally accessible \cite{doi:10.1021/nl9039636}. Further increasing $|{\cal{E}}_z|$, the system  becomes an insulator again [see Fig.  \ref{Fig. 3}(c)]. 
When  ${\cal{E}}_z=0$,  the $C_{3z}$ eigenvalues of the four low-energy bands 
are $W=e^{-i\pi/3}$, $W^*$, $-1$ and $-1$, respectively [see Fig.  \ref{Fig. 3}(c)].
However, when  ${\cal{E}}_z>{\cal{E}}_c$ (${\cal{E}}_z<-{\cal{E}}_c$), a band inversion between  the  conduction band with $C_{3z}=W$ ($C_{3z}=W^*$) and   the  valence band with $C_{3z}=-1$ occurs.
According to the theory of symmetry index \cite{fang2012bulk},  this kind of band inversion indicates that the monolayer $\rm BaMn_2S_3$  has changes from a trivial insulator to a Chern insulator with Chern number ${\cal{C}}=\pm 1$ when  ${\cal{E}}_z>{\cal{E}}_c$ (${\cal{E}}_z<-{\cal{E}}_c$).
Consequently, the Hall conductivity of system changes from $0$ to $\pm e^2/h$, depending on the direction of ${\cal{E}}_z$, and can form three Hall conductivity plateaus by varying ${\cal{E}}_z$, as shown in Fig.  \ref{Fig. 3}(d). 
This electric generation and  control of Hall conductivity may serve as a quantized version of EHE and the electric  counterpart of quantum Hall effect.

Similar to quantum Hall effect, the quantized Hall conductivity in QEHE results from finite Chern number of the system and is manifested in the appearance of the chiral edge state \cite{cage2012quantum}.
The spectrum of the (10) edge state with ${\cal{E}}_z=0, \pm 0.2$ eV/\AA ~is plotted as background of Fig. ~\ref{Fig. 3}(d), where the number and chirality  of the chiral edge state are  consistent with above analysis.

\textcolor{blue}{\emph{Conclusions.–}}
In this work, we propose a new type of Hall effect: the EHE in a two-dimensional magnetic system. By symmetry analysis, we show that this previously overlooked effect can exist in a variety of magnetic materials, including ferromagnetic, antiferromagnetic and altermagnetic materials. We also establish an analytical expression for the intrinsic EHE coefficient, and predict specific material candidate  for  EHE. In addition, based on monolayer $\rm BaMn_2S_3$, we propose a quantized version of EHE, namely QEHE. Our work not only broadens the concept of Hall effect, but also provides a new mechanism for   generating and manipulating  Hall current by electric methods.

\bibliography{ref}

\end{document}


\title{Supplemental material for ``Electric Hall Effect and Quantum Electric Hall Effect"}
\author{Chaoxi Cui}
\affiliation{Centre for Quantum Physics, Key Laboratory of Advanced Optoelectronic
Quantum Architecture and Measurement (MOE), School of Physics, Beijing
Institute of Technology, Beijing 100081, China}
\affiliation{Beijing Key Lab of Nanophotonics \& Ultrafine Optoelectronic Systems,
School of Physics, Beijing Institute of Technology, Beijing 100081,
China}

\author{Run-Wu Zhang}
\email{zhangrunwu@bit.edu.cn}
\affiliation{Centre for Quantum Physics, Key Laboratory of Advanced Optoelectronic
Quantum Architecture and Measurement (MOE), School of Physics, Beijing
Institute of Technology, Beijing 100081, China}
\affiliation{Beijing Key Lab of Nanophotonics \& Ultrafine Optoelectronic Systems,
School of Physics, Beijing Institute of Technology, Beijing 100081,
China}

\author{Yilin Han}
\affiliation{Centre for Quantum Physics, Key Laboratory of Advanced Optoelectronic
Quantum Architecture and Measurement (MOE), School of Physics, Beijing
Institute of Technology, Beijing 100081, China}
\affiliation{Beijing Key Lab of Nanophotonics \& Ultrafine Optoelectronic Systems, School of Physics, Beijing Institute of Technology, Beijing 100081,
China}

\author{Zhi-Ming Yu}
\email{zhiming\_yu@bit.edu.cn}
\affiliation{Centre for Quantum Physics, Key Laboratory of Advanced Optoelectronic Quantum Architecture and Measurement (MOE), School of Physics, Beijing
Institute of Technology, Beijing 100081, China}
\affiliation{Beijing Key Lab of Nanophotonics \& Ultrafine Optoelectronic Systems, School of Physics, Beijing Institute of Technology, Beijing 100081, China}

\author{Yugui Yao}
\email{ygyao@bit.edu.cn}
\affiliation{Centre for Quantum Physics, Key Laboratory of Advanced Optoelectronic Quantum Architecture and Measurement (MOE), School of Physics, Beijing
Institute of Technology, Beijing 100081, China}
\affiliation{Beijing Key Lab of Nanophotonics \& Ultrafine Optoelectronic Systems, School of Physics, Beijing Institute of Technology, Beijing 100081, China}

\maketitle

\section{Methods}
In this work, our results including geometry optimization and electronic structure calculations are based on density functional theory (DFT) as implemented in the Vienna Ab initio Simulation Package (VASP)~\cite{1}, using the projector augmented-wave method~\cite{2}. The exchange-correlation function is computed using the generalized gradient approximation with the Perdew-Burke-Ernzerhof realization. 
The plane-wave cutoff energy is established at 550 eV, and $\varGamma$-centered \textit{k}-mesh with sizes $21\times21\times1$ and $9\times9\times1$ are adopted for the monolayer Ca(FeN)$_2$ and monolayer BaMn$_2$S$_3$.
The crystal structure was optimized until the forces on the ions are less than 0.001 eV/\AA. A vacuum space (20 \AA) is included to avoid interactions between neighboring slabs. Since transition-metal elements have important correlation effects, we adopt the DFT+\textit{U} method~\cite{3},  testing several Hubbard \textit{U} values (\textit{e.g.}, 3 eV, 3.5 eV and 4 eV), which produced qualitatively similar
results. We showcase the main results with a \textit{U} value of Fe (Mn) set at 3.5 eV. Moreover, the maximally localized Wannier functions were constructed using the WANNIER90 code~\cite{4}.  The edge states and Chern numbers are implemented in the WannierTools package~\cite{5} based on the Green’s functions method.

\section{concrete candidate materials}
After introducing the concepts of electric Hall effect (EHE) and quantum electric Hall effect (QEHE), we illustrate our main ideas with two concrete candidate materials. In this work, we focus on layered alkaline-earth metal-decorated transition metal (TM) nitrides, \textit{viz}. Ca(FeN)$_2$ (mp-1423472), and a derivative of Ba$_2$Mg(VS$_3$)$_2$ (mp-2217817), \textit{viz}. Ba$_2$Mg(MnS$_3$)$_2$, which are proposed in the Materials Projection (https://materialsproject.org).

\begin{figure}[h]
	\centering
	\includegraphics[width=0.6\columnwidth]{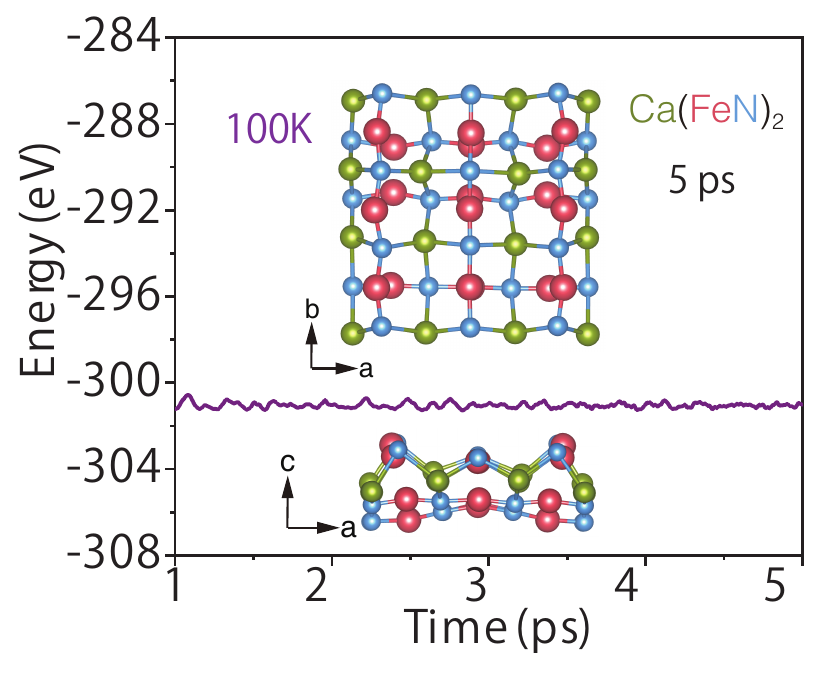}
	\caption
	{(Color online) As a typical example of testing thermal stability, herein, the total energies with AIMD simulation of the monolayer Ca(FeN)$_2$ are calculated under 100 K. Thereinto, the snapshots of the crystal structures after 5 ps in the insets.}
	\label{Fig.S1}
\end{figure}

We take monlayer Ca(FeN)$_2$ as an example to verify the feasibility of the material.
For further checking the stability of monolayer Ca(FeN)$_2$, thermal stability is examined using ab initio molecular dynamics (AIMD) with $3\times3$ supercells at 100 K. The snapshots of structure for monolayer Ca(FeN)$_2$ at 100 K is shown in Fig.~\ref{Fig.S1}. After running for 5 ps, no bond is obviously broken, although each lattice exhibited various degrees of distortion. Overall, the result suggests that monolayer Ca(FeN)$_2$ can maintain good thermal stability.

\begin{figure}[h]
	\centering
	\includegraphics[width=0.8\columnwidth]{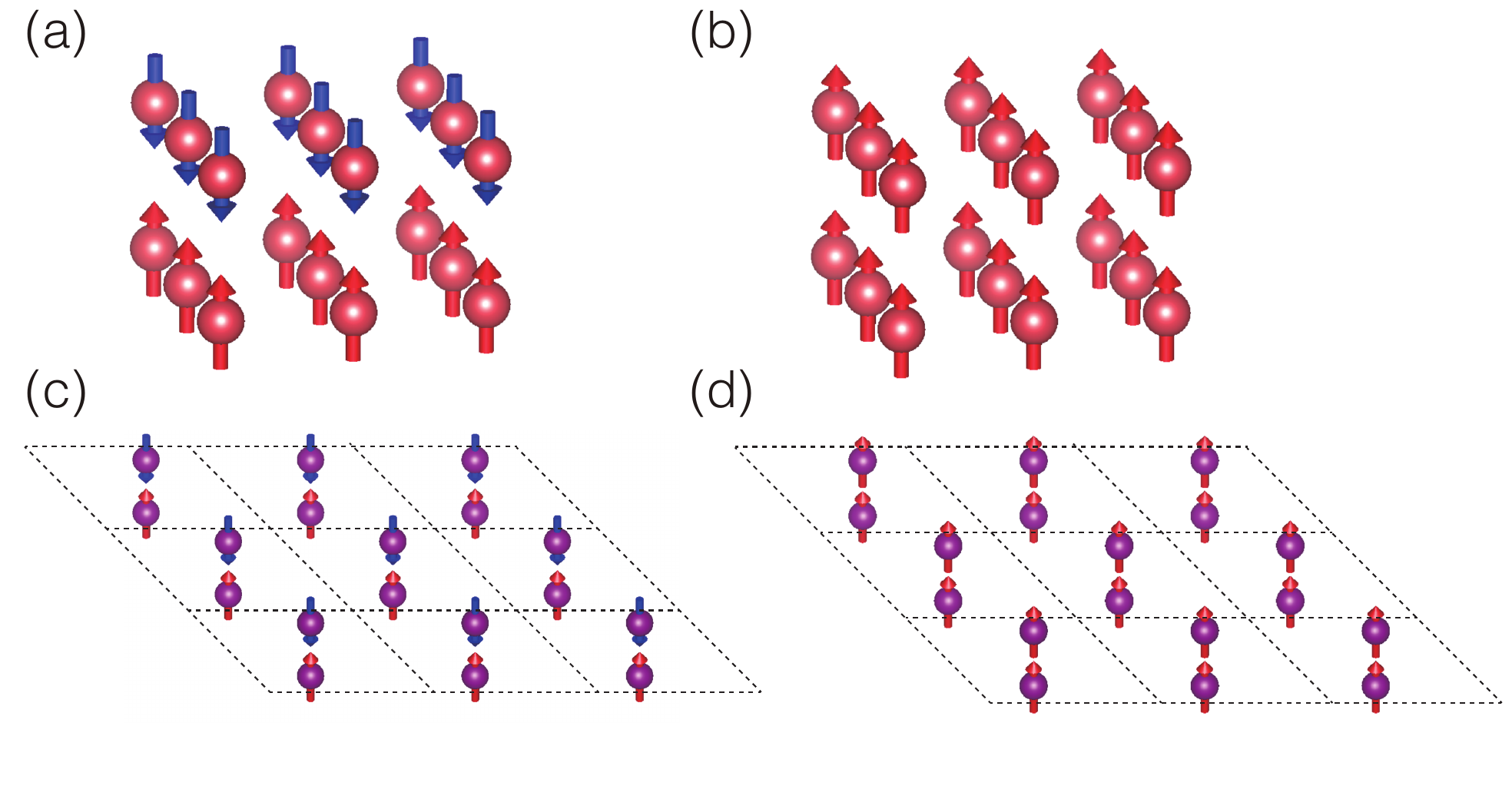}
	\caption
	{(Color online) To better check magnetic ground states of monolayer Ca(FeN)$_2$ and monolayer BaMn$_2$S$_3$, we examined two different types of feasible magnetic orders, including (a) and (c) antiferromagnetic N\'eel order (AFM-N\'eel), (b)and (d) ferromagnetic (FM). Here, the magnetic structures are examined using a $3\times3$ supercell.}
	\label{Fig.S2}
\end{figure}

To explore the intriguing physical properties of monolayer Ca(FeN)$_2$ and monolayer BaMn$_2$S$_3$, understanding their magnetic ground states is essential. Here, we investigate two possible types of magnetic ordering, specifically antiferromagnetic N\'eel order (AFM-N\'eel) and ferromagnetic (FM), in accordance with the crystal structure depicted in Fig.~\ref{Fig.S2}. Our analysis demonstrates that the AFM-N\'eel state has a lower energy than the FM state, suggesting that monolayer Ca(FeN)$_2$ and monolayer BaMn$_2$S$_3$ tend toward AFM ground states (referred to Table~\ref{Table1} and Table~\ref{Table2} for details). Moreover, we find that the AFM-N\'eel ground state with out-of-plane spin direction in Ca(FeN)$_2$ remains stable with variations in \textit{U} from 3.5 eV up to 4 eV. Calculations of magnetocrystalline anisotropy energy (MAE) at a \textit{U} value of 3 eV indicates that Ca(FeN)$_2$ behaves as a soft altermagnet, with a maximum MAE value of 0.131 meV per unit cell. Regarding monolayer BaMn$_2$S$_3$, it is observed that the AFM-N\'eel ground state with out-of-plane spin orientation in monolayer BaMn$_2$S$_3$ maintains stability with \textit{U} value 3.5 eV. MAE calculations with a \textit{U} value of 3 eV shows that monolayer BaMn$_2$S$_3$ also displays property of a soft altermagnet, with a small MAE value of 0.036 meV per unit cell. For soft altermagnet case, the magnetic characteristics of monolayer Ca(FeN)$_2$ and monolayer BaMn$_2$S$_3$ can be manipulated by changing the magnetization direction from out-of-plane to in-plane.

{\renewcommand{\arraystretch}{1.4}
	\begin{table*}
		\caption{Considering different Hubbard \textit{U} values, the relative energies (in unit of eV/Fe atom) for (a) A-type AFM, (b) FM orders in the supercell of Ca(FeN)$_2$ are listed in the following table. The $\Delta E_{MAE}$ (in unit of meV/unit cell) is defined as the difference of total energy between in-plane and out-of-plane spin directions, \textit{i}.\textit{e}., $\Delta E_{MAE}=E(\theta=90^\circ,\ensuremath{\phi}=0^\circ)-E(\theta=0^\circ,\ensuremath{\phi}=0^\circ)$. Here, $E(\theta,\phi)$ is the total energy when the spin points to the direction labeled by the polar ($\theta$) and azimuthal ($\phi$) angles. The positive (negative) value indicates a favored out-of-plane (in-plane) magnetization.} \label{Table1}
		\begin{ruledtabular}
			\begin{tabular}{cccc}
				Hubbard \textit{U} value (eV) & AFM  & FM & $\Delta E_{MAE}$ \tabularnewline
				\hline
				3 & 0 & 0.00595 & -0.131\\
				3.5 & 0 & 0.00834 & 4.217\\
				4 & 0 & 0.01204 & 2.706\\
			\end{tabular}
		\end{ruledtabular}
		
	\end{table*}
}

\begin{figure}[h]
	\centering
	\includegraphics[width=0.8\columnwidth]{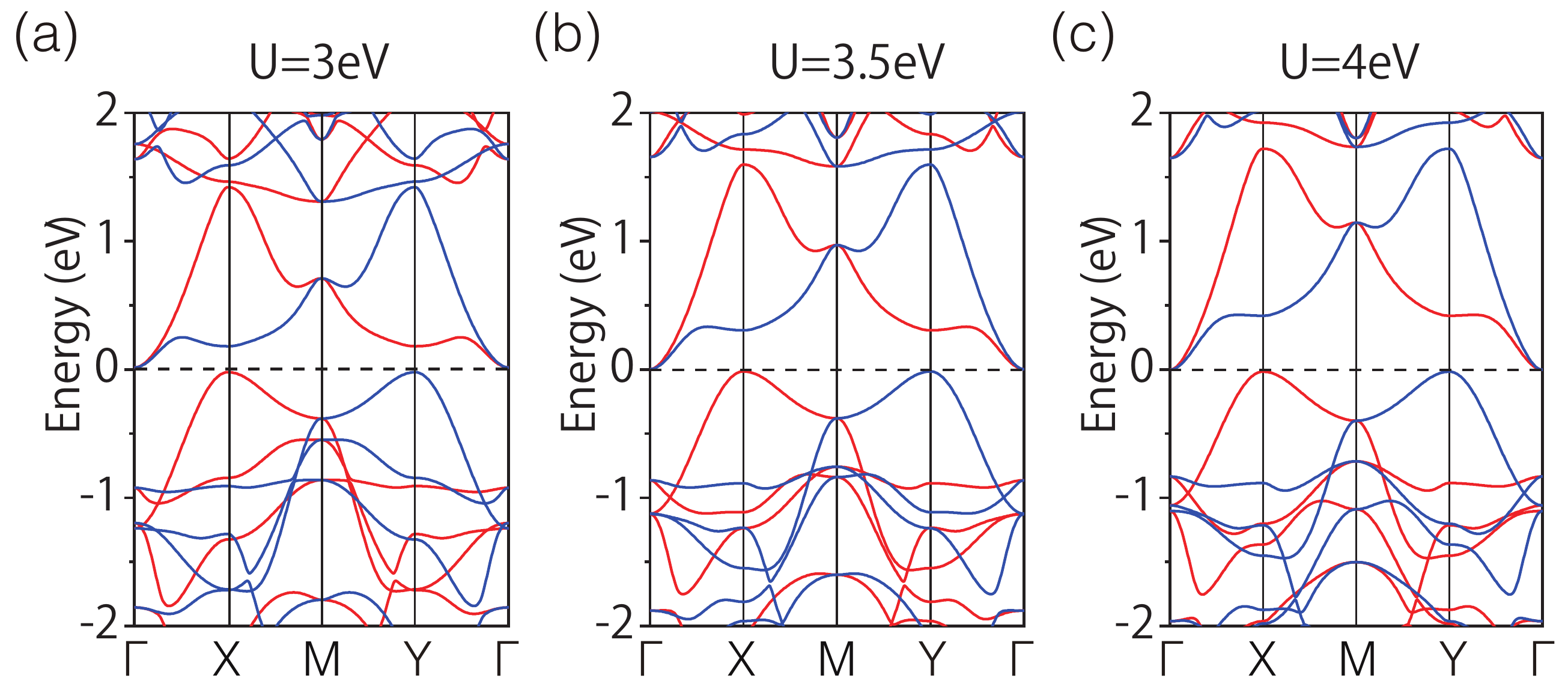}
	\caption
	{(Color online) Spin-resolved band structures of monolayer Ca(FeN)$_2$ in absence of SOC effect with different \textit{U} values, including \textit{U}=3 eV (a), \textit{U}=3.5 eV (b), and \textit{U}=4 eV (c). Red and blue lines represent spin-up and spin-down channels, respectively.}
	\label{Fig.S3}
\end{figure}

\begin{figure}[h]
	\centering
	\includegraphics[width=0.6\columnwidth]{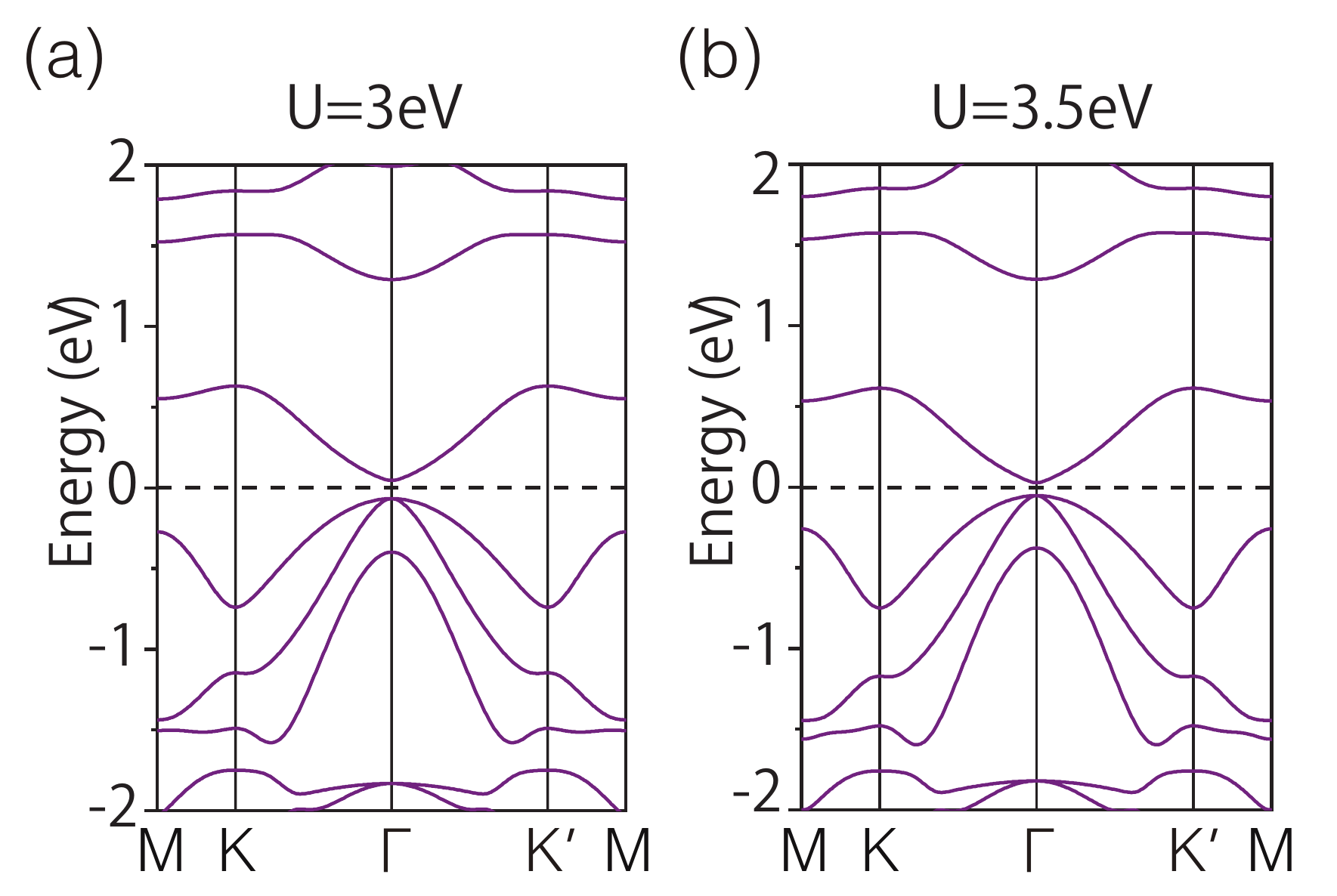}
	\caption
	{(Color online) Band structures of monolayer BaMn$_2$S$_3$ in absence of SOC effect with different \textit{U} values, including \textit{U}=3 eV (a), \textit{U}=3.5 eV (b).}
	\label{Fig.S4}
\end{figure}

{\renewcommand{\arraystretch}{1.4}
	\begin{table*}
		\caption{Considering different Hubbard \textit{U} values, the relative energies (in unit of eV/Mn atom) for (a) A-type AFM, (b) FM orders in the supercell of BaMn$_2$S$_3$ are listed in the following table. The $\Delta E_{MAE}$ (in unit of meV/unit cell) is defined as the difference of total energy between in-plane and out-of-plane spin directions, \textit{i}.\textit{e}., $\Delta E_{MAE}=E(\theta=90^\circ,\ensuremath{\phi}=0^\circ)-E(\theta=0^\circ,\ensuremath{\phi}=0^\circ)$. Here, $E(\theta,\phi)$ is the total energy when the spin points to the direction labeled by the polar ($\theta$) and azimuthal ($\phi$) angles. The positive (negative) value indicates a favored out-of-plane (in-plane) magnetization.}  \label{Table2}
		\begin{ruledtabular}
			\begin{tabular}{cccc}
				Hubbard \textit{U} value (eV) & AFM  & FM & $\Delta E_{MAE}$ \tabularnewline
				\hline
				3 & 0 & 0.03206 & -0.036\\
				3.5 & 0 & 0.01482 & 0.044\\
			\end{tabular}
		\end{ruledtabular}
		
	\end{table*}
}
Given that the candidate materials studied all contain transition metal (TM) atoms with strongly correlated 3\textit{d} electrons, we adopt Hubbard \textit{U} with moderate values for 3\textit{d} TM atoms to check our results. To fully validate our ideas, we test several \textit{U} values including 3 eV, 3.5 eV and 4 eV, as shown in Fig.~\ref{Fig.S3} and Fig.~\ref{Fig.S4}. Notably, the key band features remain robust over a wide range of Hubbard \textit{U} values. Increasing the \textit{U} value mainly resulted in increased band gaps.

\section{List of electric Hall effect compatible magnetic point (layer) groups}
Based on the symmetry requirements detailed in Table I, we have identified all magnetic point groups (MPGs) and magnetic layer groups (MLGs) that are compatible with the electric Hall effect, as shown in Table~\ref{Table3}. This identification significantly simplifies the search for suitable materials and provides valuable guidance for classifying potentially feasible materials. We have determined that 62 of the 528 MLGs are compatible with the electric Hall effect.

{\renewcommand{\arraystretch}{1.4}
	\begin{table*}
		\caption{List of magnetic point groups and the magnetic layer group that permit EHE. The symmetries belong to $M_{z}\mathcal{TS}$ which is critical for EHE , are presented in second column.}  \label{Table3}
		\begin{ruledtabular}
			\begin{tabular}[t]{lll|lll}
				MPG & $M_{z}\mathcal{TS}$ & MLG & MPG & $M_{z}\mathcal{TS}$ & MLG\tabularnewline
				$\bar{1^{'}}$ & $P\mathcal{T}$ & 2.3.6 & $4/m'$ & $S_{4z}\mathcal{T},\mathcal{PT}$ & 51.3.364,52.3.371 \\
				$m'$ & $M_{z}\mathcal{T}$ & 4.3.14 5.3.19  & $422$ & $C_{2\|}$ & 53.1.374,54.1.381 \\
				{$2/m'$} & {$P\mathcal{T}$} & 6.4.24,7.4.31,14.3.68,15.3.78, &
				$\bar{4}'2m'$,$\bar{4}'m'2$ & $S_{4z}\mathcal{T},C_{2\|}$ & 57.4.401,58.4.408 ,59.5.414,60.5.421 \\
				&  & 16.3.85,17.3.92,18.3.97  & $4/m'm'm'$ & $S_{4z}\mathcal{T},C_{2\|},\mathcal{PT}$ & 61.3.424,62.3.437,63.3.446,64.3.455 \\
				$2$ & $C_{2\|}$ & 8.1.34,9.1.41,10.1.45  & $\bar{3^{'}}$ & $S_{3z}\mathcal{T},P\mathcal{T}$ & 66.3.466 \\
				$222$ & $C_{2\|}$ & 19.1.104,20.1.111,21.1.118,22.1.122  & $321$ & $C_{2\|}$ & 68.1.470 \\
				{$m'2m'$} & {$C_{2\|},M_{z}\mathcal{T}$} & 27.3.156,28.3.169,29.3.176,30.3.183 & $\bar{3^{'}}1m'$,$\bar{3^{'}}m'1$ & $S_{3z}\mathcal{T},C_{2\|},\mathcal{PT}$ & 71.3.481 ,72.3.486 \\
				&  & 31.3.190,32.3.199,33.3.204 & $\bar{6^{'}}$ & $S_{6z}\mathcal{T},M_{z}\mathcal{T}$ & 74.3.494 \\
				&  & 34.3.209,35.3.214,36.3.223 & $6/m'$ & $S_{6z}\mathcal{T},M_{z}\mathcal{T},P\mathcal{T}$ & 75.3.497 \\
				{$m'm'm'$} & {$C_{2\|},\mathcal{PT}$} & 37.3.232,38.3.245,39.3.258,40.3.265  & $622$ & $C_{2\|}$ & 76.1.500 \\
				&  & 41.3.278,42.3.291,43.3.300,44.3.309 & $\bar{6^{'}}m'2$,$\bar{6^{'}}2m'$ & $S_{6z}\mathcal{T},M_{z}\mathcal{T},C_{2\|}$ & 78.5.514 ,79.4.518\\
				&  & 45.3.316,46.3.325,47.3.332,48.3.345 & $6/m'm'm'$ & $S_{3z}\mathcal{T},C_{2\|},\mathcal{PT},M_{z}\mathcal{T}$ & 80.3.522 \\
				$\bar{4^{'}}$ & $S_{4z}\mathcal{T}$ & 50.3.360  &  &  & \\
			\end{tabular}
		\end{ruledtabular}
		
	\end{table*}
}

\section{Derivation of Eq. (7) in main text}
When ${\mathcal{E}}_z$ is  finite but weak, the   Hamiltonian of the perturbed system  reads 
\begin{equation}
	\hat{H}=\hat{H}_{0}+g_{E}\mathcal{E}_z\hat{z}.
\end{equation}
According to the  standard perturbation theory \cite{fernandez2000introduction}, the ${\mathcal{E}}_z$ has a linear correction on both band dispersion and wave function  of system, expressed as 
\begin{align}{\label{pertubed eu}}
	\tilde{\epsilon}_n&=\epsilon_n+P_{nn}\mathcal{E}_z, \\ 
	\left|\tilde{u}_{n}\right\rangle &=\left|u_{n}\right\rangle +\mathcal{E}_z\sum_{l\ne n}\frac{P_{ln}}{\epsilon_{n}-\epsilon_{l}}\left|u_{l}\right\rangle, 
\end{align}
where $P_{nm}=g_E \langle u_{n}(\boldsymbol{k})|\hat{z}|u_{m}(\boldsymbol{k})\rangle$ and $u_{n}(\boldsymbol{k})$ is unperturbed wave function. The intrinsic Hall conductivity of the perturbed system  is 
\begin{equation}
	\sigma_{x y}=\frac{e^{2}}{\hbar} \sum_{n}\int \frac{d^2 k}{(2 \pi)^2} 
	f(\tilde{\varepsilon}_{n,k})\tilde{\Omega}_n(\boldsymbol{k}),
\end{equation}
where the perturbed  Fermi-Dirac distribution $f(\tilde{\varepsilon}_{n}) $ can be easily obtained as 
\begin{eqnarray}
	f(\tilde{\varepsilon}_{n}) & = & f(\varepsilon_{n})+\frac{\partial f(\varepsilon_n)}{\partial\varepsilon_n}P_{nn}(\boldsymbol{k})\mathcal{E}_z.
\end{eqnarray}
We then calculate the  perturbed Berry curvature   $\tilde{\Omega}_n(\boldsymbol{k})$. The expression of $\tilde{\Omega}_n(\boldsymbol{k})$ is 
\begin{eqnarray}{\label{pertubed Berry}}
\tilde{\Omega}_{n}(k)=-2\textrm{Im}\sum_{m\neq n}\frac{\left\langle \tilde{u}_{n}\left|\hat{v}_{x}\right|\tilde{u}_{m}\right\rangle \left\langle \tilde{u}_{m}\left|\hat{v}_{y}\right|\tilde{u}_{n}\right\rangle }{\left(\tilde{\epsilon}_{m}-\tilde{\epsilon}_{n}\right)^{2}},
\end{eqnarray}
where velocity element of the perturbed system to the first order of ${\mathcal{E}}_z$ is 
\begin{eqnarray}
	\left\langle \tilde{u}_{n}\left|\hat{v}_{x(y)}\right|\tilde{u}_{m}\right\rangle &=&\left\langle u_{n}\right|\hat{v}_{x(y)}\left|u_{m}\right\rangle +\mathcal{E}_z\sum_{l\neq n}\frac{P_{ln}^{*}}{\varepsilon_{n}-\varepsilon_{l}}\left\langle u_{l}\right|\hat{v}_{x(y)}\left|u_{m}\right\rangle +\mathcal{E}_z\sum_{l\neq m}\frac{P_{lm}}{\varepsilon_{m}-\varepsilon_{l}}\left\langle u_{n}\right|\hat{v}_{x(y)}\left|u_{l}\right\rangle. 
\end{eqnarray}
Then, each term in the perturbed Berry curvature  is obtained as 
\begin{eqnarray}
	\frac{\left\langle \tilde{u}_{n}\left|\hat{v}_{x}\right|\tilde{u}_{m}\right\rangle \left\langle \tilde{u}_{m}\left|\hat{v}_{y}\right|\tilde{u}_{n}\right\rangle }{\left(\tilde{\epsilon}_{m}-\tilde{\epsilon}_{n}\right)^{2}}	&=&	\frac{v_{x}^{nm}v_{y}^{mn}}{\left(\varepsilon_{m}-\varepsilon_{n}\right)^{2}}-2\mathcal{E}_z\frac{P_{mm}-P_{nn}}{\left(\varepsilon_{m}-\varepsilon_{n}\right)^{3}}v_{x}^{nm}v_{y}^{mn} \\ \notag 
	&+&\mathcal{E}_{z}\sum_{l\ne m}\frac{P_{lm}v_{x}^{nl}v_{y}^{mn}+P_{lm}^{*}v_{x}^{nm}v_{y}^{ln}}{\left(\epsilon_{n}-\epsilon_{m}\right)^{2}\left(\epsilon_{m}-\epsilon_{l}\right)}+\mathcal{E}_{z}\sum_{l\ne n}\frac{P_{ln}^{*}v_{x}^{lm}v_{y}^{mn}+P_{ln}v_{x}^{nm}v_{y}^{ml}}{\left(\epsilon_{n}-\epsilon_{m}\right)^{2}\left(\epsilon_{n}-\epsilon_{l}\right)}
\end{eqnarray}
where  only the terms up to the first order of ${\mathcal{E}}_z$ are kept.
Consequently, we have 
\begin{eqnarray}
	\tilde{\Omega}_{n}(\boldsymbol{k}) & = & \Omega_{n}(\boldsymbol{k})+\Lambda_n(\boldsymbol{k})\mathcal{E}_z,
\end{eqnarray} 
with $\Omega_{n}(\boldsymbol{k})=-2\textrm{Im}\sum_{m\neq n}\frac{v_{x}^{nm}v_{y}^{mn}}{\left(\varepsilon_{m}-\varepsilon_{n}\right)^{2}}$, and 
\begin{eqnarray}
	\Lambda_n(\boldsymbol{k}) &=& -2\textrm{Im}\sum_{m\neq n}\left[-2\frac{P_{mm}-P_{nn}}{\left(\varepsilon_{m}-\varepsilon_{n}\right)^{3}}v_{x}^{nm}v_{y}^{mn} +\sum_{l\neq m}\frac{P_{lm}^{*}v_{y}^{ln}v_{x}^{nm}+P_{lm}v_{x}^{nl}v_{y}^{mn}}{\left(\varepsilon_{n}-\varepsilon_{m}\right)^{2}\left(\varepsilon_{m}-\varepsilon_{l}\right)} 
	+\sum_{l\neq n}\frac{P_{ln}v_{y}^{ml}v_{x}^{nm}+P_{ln}^{*}v_{x}^{lm}v_{y}^{mn}}{\left(\varepsilon_{n}-\varepsilon_{m}\right)^{2}\left(\varepsilon_{n}-\varepsilon_{l}\right)}\right].\notag\\
\end{eqnarray} 

Substituting $f(\tilde{\varepsilon}_{n})$ and $\tilde{\Omega}_{n}(\boldsymbol{k})$ into $\sigma_{x y}$ and keeping the terms  up to the first order of ${\mathcal{E}}_z$, we have 
\begin{equation}
	\sigma_{x y}=\sigma_{x y}^0+{\mathcal{E}}_z \frac{e^{2}}{\hbar} \sum_{n}\int \frac{d^2 k}{(2 \pi)^2}\left[\frac{\partial f(\varepsilon_n)}{\partial \varepsilon_n}  P_{nn}(\boldsymbol{k}) \Omega_n(\boldsymbol{k}) +f(\varepsilon_{n}) \Lambda_n(\boldsymbol{k})\right],
\end{equation}
with 
\begin{equation}
	\sigma_{x y}^0=\frac{e^{2}}{\hbar} \sum_{n}\int \frac{d^2 k}{(2 \pi)^2}f(\varepsilon_{n,k})\Omega_n(\boldsymbol{k})=0,
\end{equation}
as the original system has vanishing intrinsic Hall conductivity.
Finally, according to the equation $\sigma_{x y}= \chi_{xy}{\mathcal{E}}_z$, we obtain the expression of the  intrinsic EHE coefficient  $\chi_{xy}$ as  
\begin{equation}
	\chi_{xy}=\frac{e^2}{\hbar} \sum_n \int  \frac{d^2 k}{(2 \pi)^2}\left[\frac{\partial f(\varepsilon_n)}{\partial \varepsilon_n}  P_{nn}(\boldsymbol{k}) \Omega_n(\boldsymbol{k}) +f(\varepsilon_{n}) \Lambda_n(\boldsymbol{k})\right],
\end{equation}
which is exactly the Eq.  (7) in main text.

\section{Effective $\mathcal{T}$  symmetry in  collinear or coplanar magnetic materials}
When spin degree of freedom is considered but the  SOC effect is ignored, the symmetry of the  systems should be described  by spin space group \cite{brinkman1966theory,litvin1974spin,liu2022spin,xiao2023spin,ren2023enumeration,jiang2023enumeration}, the operator of which  is generally  written as
\begin{align}
	g=\left\{U \| R | \boldsymbol{\tau}\right\},
\end{align}
where $U$ acts on spin,  $R$ acts on lattice and $\boldsymbol{\tau}$ is translation. 
For collinear or coplanar magnetic materials, they all have  an operator:  $\{C_{2\bot}\mathcal{T}\|E|0\}$. Here, $C_{2\bot}$ stands for a 180 degree rotation along axis normal to the direction of all spins  in the system. 
Since $\{C_{2\bot}\mathcal{T}\|E|0\}$ does not exchange  the spin index and the $C_{2\bot}$ only acts on spin, each spin channel of the system should  have an effective  $\mathcal{T}$ symmetry.
Thus,  there will not be either anomalous Hall effect or intrinsic EHE in collinear and coplanar magnetic materials without SOC.

\bibliography{supref}